# Electrical resistivity and tunneling anomalies in CeCuAs$_2$

E.V. Sampathkumaran[a,*], T. Ekino[b], R.A. Ribeiro[b], Kausik Sengupta[a], T. Nakano[c], M. Hedo[c], N. Fujiwara[c], M. Abliz[c], Y. Uwatoko[c], S. Rayaprol[a], Th. Doert[d], J.P.F. Jemetio[d]

[a]*Tata Institute of Fundamental Research, Homi Bhabha Road, Colaba, Mumbai – 400005, India*

[b]*Faculty of Integrated Arts and Sciences, Hiroshima University, Higashi-Hiroshima 739-8521*

[c]*ISSP, University of Tokyo, Kashiwa, Chiba 2778581, Japan*

[d]*Technische Universität Dresden, Institut für Anorganishe Chemie, D-01062, Dresden, Germany*

**Abstract**

The compound CeCuAs$_2$ is found to exhibit negative temperature (T) coefficient of electrical resistivity ($\rho$) under ambient pressure conditions in the entire T-range of investigation (45 mK to 300 K), even in the presence of high magnetic fields. Preliminary tunneling spectroscopic measurements indicate the existence of a psuedo-gap at least at low temperatures, thereby implying that this compound could be classified as a Kondo semi-conductor, though $\rho$(T) interestingly is not found to be of an activated type

Following the observation [1,2] of interesting electrical resistivity ($\rho$) anomalies for many rare-earths (R) (other than Ce) in the series RCuAs$_2$, crystallizing in ZrCuSi$_2$-type tetragonal structure, we have taken up the Ce compound, CeCuAs$_2$, for investigations. The results of initial investigations have been reported in Ref. 3. Subsequently, we have carried out further investigations, viz., high-field/high-pressure $\rho$ studies, as well as preliminary tunneling spectroscopic experiments on this compound. We briefly present these findings and it appears that this compound could be classified as a Kondo semi-conductor at least at low temperatures.

Some of the experimental details including sample preparation have been mentioned in Ref. 3. The $\rho$ data under ambient pressure conditions, reported in figure 1, were performed (T= 1.8 to 300 K in the present study) on freshly prepared specimens with a Physical Property Measurement System (Quantum Design). In addition, we report the $\rho$(T) behavior (2.5-300 K) at a high pressure (10GPa) obtained from a cubic-anvil pressure apparatus [4]. Preliminary tunneling conductance experiments at 4.2, 18.2 and 21.8 K were performed by a break-junction technique [5].

---

[*] Corresponding author. Tel.: 0091 22 2280 4545; fax: 0091 22 2280 4610; e-mail: sampath@tifr.res.in.

The values of ρ for the freshly prepared specimen fall in the range of several mΩcm in agreement with our previous findings [3]. In figure 1, we also show the ρ(T) behavior of non-magnetic analogues. It is obvious that the Ce compound exhibits negative dρ/dT in the entire T-range investigation (measured down to 45 mK in Ref. 3) under ambient pressure conditions, increasing by a factor of about 2 as T is lowered from 300 K to 45 mK. Considering that the non-magnetic compounds exhibit positive temperature coefficient of ρ, the observed negative dρ/dT must be 4f-related. The question therefore arises whether this behavior of ρ for R= Ce is due to the development of a pseudo-gap or whether it is due to Kondo behavior.

Considering [3] that (i) no activated behavior could be seen in the entire T-range of investigation, but ρ varies logarithmically with T above 30 K and (ii) the thermopower above 5 K is small, rather mimicking the behavior of metallic, trivalent, Ce-based Kondo lattices, we are tempted [3] to believe that the negative dρ/dT may arise from the Kondo effect at least in the high T-range (above 30 K), though at lower temperatures, the question was left open.

This open question could be addressed with the results of preliminary tunneling conductance experiments (see figure 2). It is obvious that dI/dV exhibits well-defined symmetric shoulders with respect to bias, typical of gap effects. However, in contrast to other Kondo semi-conductors [5], we see two values for the peak-to-peak gap-edge separation (which is equivalent to twice of gap-width), ±150meV and 500 meV, say at 4.2 K, with the former gap structure (150 meV) disappearing at 21.8 K. While the double-gap feature is fascinating, these tunneling experiments seem to provide evidence for the pseudo-gap behavior of $CeCuAs_2$ even around 20 K, though an indication of a psuedo-gap in the heat-capacity data is visible only below 5 K [3].

We have performed magnetoresistance (MR= [ρ(H)-ρ(0)]/ρ(0)) measurements upto 140 kOe (see an inset in figure 1) down to 1.8 K. It is remarkable that dρ/dT remains negative even at fields as high as 140 kOe (see Fig. 1). However, there is a change in the magnitude of the slope of ρ(T) plot with decreasing temperature, but this change becomes non-monotonic with increasing H, particularly in the vicinity of 14 K (see inset of figure 1). There are also qualitative changes in the shapes of the ρ(H) curves as T is lowered across 14 K (compare the MR curves for 1.8 and 20 K in the inset). Considering that there is no evidence [3] for magnetic ordering above 45 mK (as evidenced by our NMR experiments [6] as well down to 0.6 K), this finding is noteworthy. It is not clear what the significance of this temperature is. While MR is negligible above about 30 K, the ρ values decrease noticeably with increasing H as the temperature is lowered, attaining a value as large as about –16% at 1.8 K for H= 140 kOe.

Finally, we stress that the temperature coefficient of ρ becomes positive at high pressures [4], as shown in figure 1 for the application of 10 GPa. This can interpreted by proposing that the proposed pseudo-gap closes with the application of pressure and/or the Kondo temperature undergoes a dramatic increase with the application of pressure.

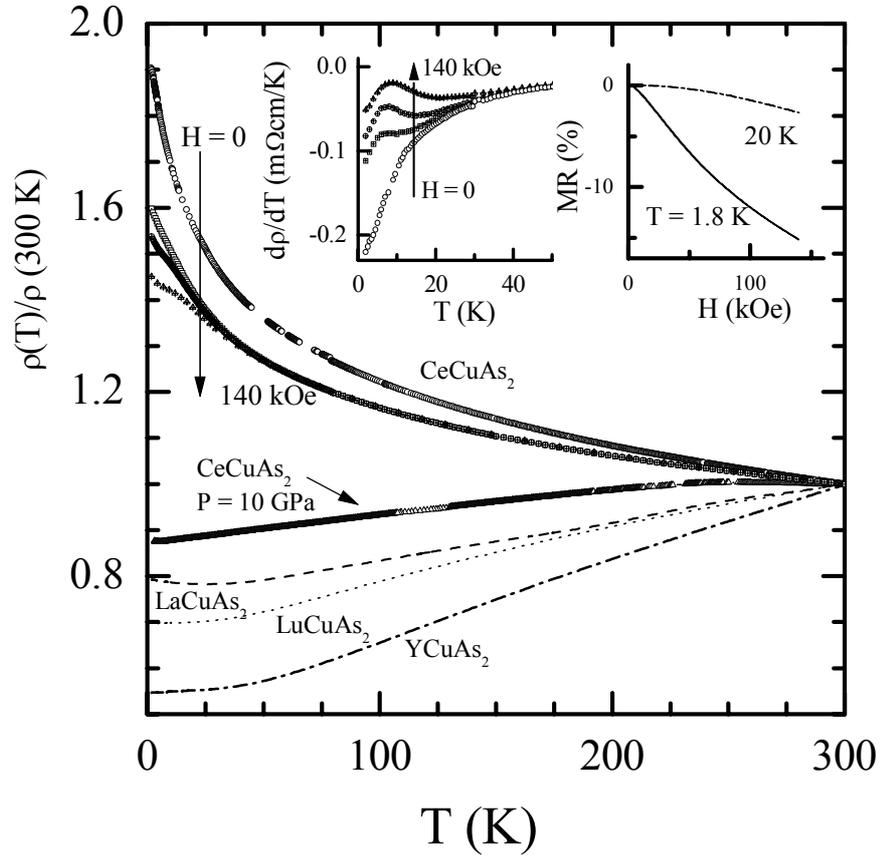

**Fig. 1**: Electrical resistivity ($\rho$) of RCuAs$_2$ (R= Ce, La, Y, Lu) as a function of temperature normalized to respective 300 K values. For R= Ce, the data in the presence of few selected magnetic fields (H= 0, 50, 80, 140 kOe increasing in the direction of arrow) well as the temperature dependence under a high pressure (10 GPa, zero field) are also shown. In the insets, the temperature derivative of $\rho$ at low temperatures for R= Ce for various H and the dependence of $\rho$ on H at 1.8 and 20 K are shown.

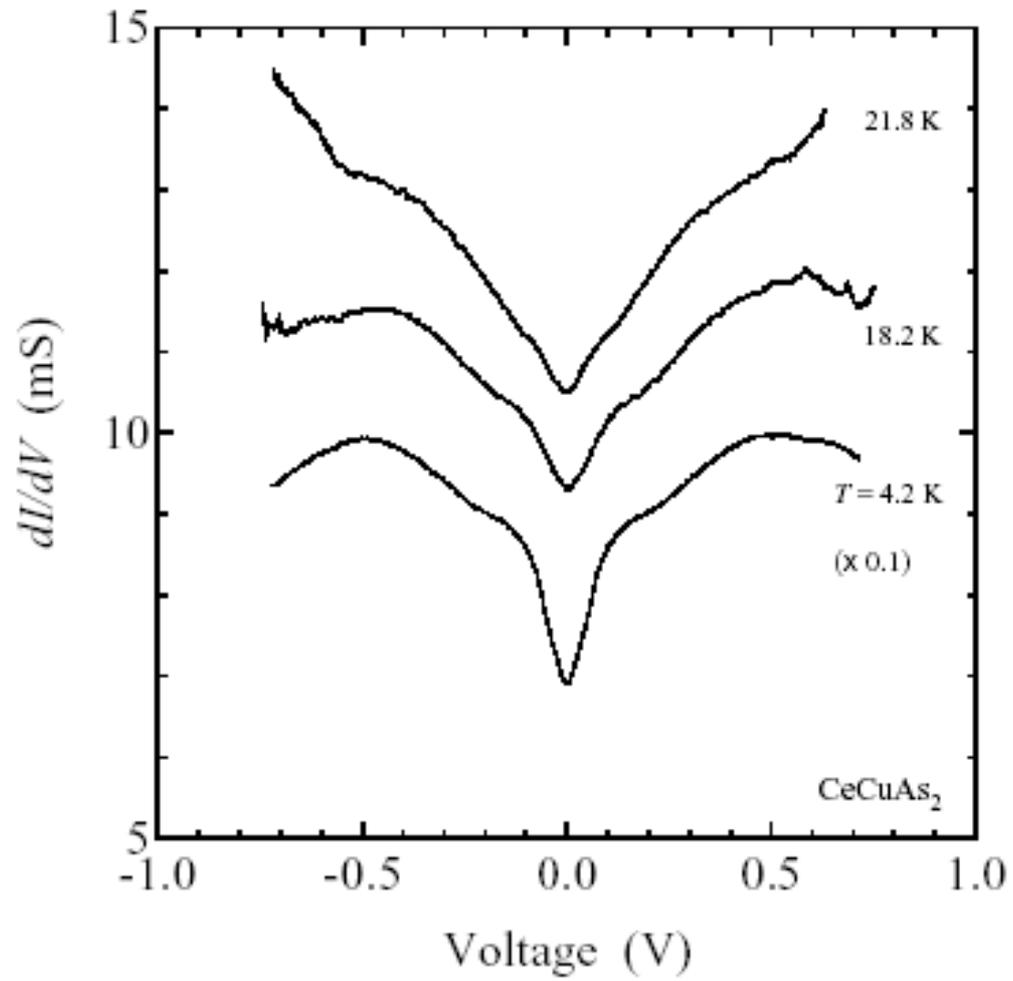

**Fig. 2**: Tunneling conductance for CeCuAs$_2$.